
\input Jnl.TEX
\input Reforder.TEX

\def\s{\sigma}
\def\D{\Delta}
\def\alp{\alpha}

\def\eps{\epsilon}
\def\lam{\lambda}
\def\del{\delta}
\def\th{\theta}
\def\dg{\dagger}

\title {Incommensurate and collinear Phases in a doped quantum
anti-ferromagnet.}

\author W. Barford and S. Jadhav
\affil {\rm Department of Physics, The University of Sheffield,
Sheffield, S3 7RH,
United Kingdom.}

\abstract

The Schwinger Boson mean field theories of the
`t-J' model are extended by the consideration of anisotropic order
parameters. This has two effects. First, a collinear phase, in
which the spins are anti-ferromagnetically aligned in one direction
and ferromagnetically aligned in the other, is found to be stable
over a significant range of the phase diagram. Second, the (1,1) and
(1,0) spiral phases become very close in energy. The inclusion of
weak intra-sublattice coupling may therefore stabilise the (1,0)
spiral with respect to the (1,1) spiral, thus harmonising theory and
experiment.

PACS numbers: 75.10.-b, 75.25.+z

\endtopmatter

\subhead{1. Introduction}

It is both experimentally and theoretically established that the
undoped high temperature superconductors are described by a two
dimensional spin 1/2 Heisenberg anti-ferromagnet. Three dimensional
coupling stabilises the long range N\'{e}el order to temperatures of
roughly 300 K. Upon doping, however, the long range order is soon
destroyed to be replaced by short range anti-ferromagetic
correlations, which are either incommensurate, as in
La$_{2-x}$Sr$_x$CuO$_4$\refto{hayden}, or commensurate,
as in YBa$_2$Cu$_3$O$_{7-\del}$, for example.

There have been many theoretical attempts to explain this
behaviour. One approach is to treat the spins semi-classically.
The hole motion then leads to a (1,1)
spiral distortion of the spin background\refto{shraiman,barford}.
Another approach is a weak coupling RPA calculation in which the most
divergent terms in the magnetic susceptibilty arise from the Fermi
surface nesting wave vectors\refto{lu}. This approach has the virtue
of explicitly building in the Fermi surface structure, and hence is
easily able to explain the discrepencies between different compounds
from their different Fermi surfaces. However, it is implicitly
assumes the underlying existence of a Fermi liquid, which is
somewhat at variance with normal state transport properties.

A further approach is to take the strong coupling limit and to use a
slave representation to explicitly prohibit double occupancy. Of
the two (formally equivalent) representations we elect to use the
Slave fermion-Schwinger boson representation in which the fermion
degree of freedom controls the occupancy. This method correctly
predicts long range anti-ferromagnetism at half filling and the
Nagaoka ferromagnet for infinitesimally small doping and infinite
coupling, so it is a natural starting point when considering doping.

The Schwinger boson method has been used previously to study the
mean field theory of the `t-J' model\refto{yoshioka, kane} (which we
define shortly). It was found that doping leads to an incommensurate
spiral along the (1,1) direction, which is in contrast to the
experimental results of Cheong {\it et al.}\refto{hayden} who found
incommensurate spirals along the (1,0) direction. The inclusion of a
next nearest neighbour hopping term, however, leads to the
stabilisation of the (1,0) spiral for some parameter
ranges\refto{kane}. The mean field theory without fluctuations also
predicts long range order for most doping values, whereas short
range correlations are observed. Later work has shown that the long
range order is destroyed by the fluctuations\refto{gan}. This theory
therefore seems a promising one for a complete explanation of the
phase diagram.

In most of the mean field analyses it is customary to assume that
the order parameters are isotropic. However, if this restriction
is relaxed new phases are found which are energetically favourable.
In particular, we find that there is a phase transition from the
(1,1) spiral to a collinear phase, in which the spins are
anti-ferromagneticaly aligned in one direction and
ferromagnetically aligned in the other, before the onset of
ferromagnetism. Furthermore, the energy of the (1,0) spiral becomes
very close to that of the (1,1) spiral. The plan of this paper
is as follows. First we introduce the `t-J' model to establish
notation. Then, in \S2, we describe the Schwinger boson mean field
method. \S3 contains our results, and we conclude in \S4.

The `t-J' model can be obtained from the
Hubbard hamiltonian by formally projecting out double occupied
states, but allowing virtual states of double occupancy. This leads
to a Heisenberg exchange term with a coupling constant,
$J=4t^2/U$. It is more convenient, however, to treat this as a
model Hamiltonian with $t$ and $J$ as indepenedent variables.
 The `t-J' model
then reads:
$$ H=
-t\sum_{<ij>\s}{\tilde c}_{i\s}^\dagger {\tilde c}_{i\s} + h.c. +
J\sum_{<ij>}\left(S_i.S_j - {1 \over 4}n_i n_j \right),
\eqno(1) $$
where ${\tilde c}_{i\s}$ acts only in the empty and singly
occupied subspace, and is defined as $(1-n_{i{\bar \s}})c_{i\s}$.
The sum, $<ij>$, is over nearest neighbours.

With the introduction of the singlet operator,
$\D_{ij} = {1 \over \sqrt 2}\sum_\s {\rm sgn}(\s)c_{i\s}c_{j{\bar
\s}}, $
eqn(1) can be rewritten as,
$$
H = -t\sum_{<ij>\s}{\tilde c}_{i\s}^\dagger {\tilde c}_{i\s} + h.c.
- J\sum_{<ij>} \D_{ij}^\dagger\D_{ij}.
\eqno(2)
$$

\subhead{2. Mean Field Analysis}

To handle the constraint of no double occupancy we adopt the slave
fermion representation whereby the physical electron operator is
factorised into a spinless fermion, $f_i$, and the SU(2)
Schwinger boson, $b_{i\s}$, $\it i.e.$
$$
c_{i\s}^\dg=f_i b_{i\s}^\dg.
$$
The fermions and bosons
necessarily satisfy the constraint  $\sum_\s b_{i\s}^\dg b_{i\s}
+f_i^\dg f_i =1$, and so we may therefore identify the $f_i^\dg$ as
creating a hole and $b_{i\s}^\dg$ as creating a spin on the $ith.$
site. Using this representation eqn(2) then becomes,
$$ H = -t \sum_{<ij>\s} f_i
f_j^\dagger b_{i\s}^\dagger b_{j\s} + h.c. - J \sum_{<ij>} \left(1 -
f_i^\dagger f_i \right)\D_{ij}^{b\dagger} \D_{ij}^b \left(1-
f_j^\dagger f_j \right), \eqno(3)
$$
where $\D_{ij}^b = {1 \over \sqrt 2}\sum_\s {\rm sgn}(\s)b_{i\s}
b_{j{\bar \s}}.$

We use the saddle point approximation to solve eqn(3), and
introduce the following mean field order parameters\refto{auerbach},
$$
Q_{ij}=\sum_\s<b_{i\s}^{\dagger}b_{j\s}>,
\eqno(4a)
$$
$$
F_{ij}=<f_i^\dagger f_j>
\eqno(4b)
$$ and
$$
D_{ij}=<\D_{ij}^b>.
\eqno(4c)
$$
$Q_{ij}$ is a measure of the ferromagnetic
alignment between neighbouring spins. This is illustrated by noting
that for classical spins represented by the unit vectors $\^{\Omega}$
it is given by\refto{barford}
$$
Q_{ij}=(1-\del) \sqrt{{1 \over 2}\left(1+\^{\Omega_i}.\^{\Omega_j}
\right)} e^{i\^{\omega}(\^{\Omega_i},\^{\Omega_j};\^{z})},
\eqno(5)
$$
where $\^{\omega}(\^{\Omega_i},\^{\Omega_j};\^{z})}$ is the solid
angle spanned by the two spins and the global axis of quantisation.
$\del$ is the hole concentration, so $(1-\del)$ is a mean field
estimate of the average local spin concentration. $F_{ij}$ is a
measure of the bosonic band width, while $D_{ij}$ is the singlet
order parameter, and hence measures the anti-ferromagnetic
alignement.  All of the order parameters are assumed to be
translationally  invariant, with Q and F being even and D being odd
under a space inversion.

In general, we are at liberty to choose the
ratio of $Q_x$ and $Q_y$ and allow $D_x$, $D_y$, $F_x$ and $F_y$ to
be determined self-consistently. Setting this ratio to unity
results in the (1,1) spiral, while setting it to zero results in the
(1,0) spiral. Alternatively, we
can choose the ratio of $D_x$ and $D_y$, while the other four order
paprameters are then self-consistently determined. In particular,
setting this ratio to zero results in the collinear phase.  The term
$J(1-f_i^{\dag} f_i)(1-f_j^{\dag} f_j)$ is replaced by its mean field
value $J_\del = J(1-\del)^2$.

By decoupling eqn(3) with these order parameters the spin and
charge degrees are freedom are decoupled and the
following Hamiltonians for the charge and spin degrees
of freedom are obtained, $$
{\tilde H}^f =  \sum_k (\eps_k^f +\lam_f)f_k^\dagger f_k
+tN(F_xQ_x+F_yQ_y)
\eqno(6a)
$$
and
$$
{\tilde H}^b= \sum_k \left( b_{k\uparrow}^\dagger , b_{-k \downarrow}
\right) \left( \matrix{
\lambda_b + M_k & \gamma_k  \cr
\gamma_k^* & \lambda_b + M_{-k} \cr}
\right)
\left(
\matrix{
b_{k\uparrow} \cr
 b_{-k \downarrow}^\dagger \cr }
\right)
\hskip 1.2 truein
$$
$$\hskip 1.2 truein -\lam_bN +J_\del N(D_x^2+D_y^2) -
J_\del(Q_x^2+Q_y^2)N/4.
\eqno(6b)
$$
Notice that (6a) and (6b) have been defined in the grand canonical
ensemble to ensure that the constraints $<f_i^\dagger f_i>=\del$ (=
hole doping) and $\sum_\s<b_{i\s}^\dagger b_{i\s}> = 1 -\del$ are
satisfied on the average. This is done via the introduction of the
fermionic and bosonic chemical potentials, $\lam_f$ and $\lam_b$,
respectively. We define $$
M_k = 2\left( \left( F_xt+ {Q_xJ_\del \over 4} \right)\cos k_x +
\left( F_yt+ {Q_yJ_\del \over 4} \right)\cos k_y \right)
\eqno(7)
$$
and
$$
\gamma_k = \sqrt{2}iJ_\del(D_x\sin k_x + D_y\sin k_y).
\eqno(8)
$$

The fermionic Hamiltonian is trivially diagonalised giving the
spectrum: $$
\eps_k^f = -2t(Q_x\cos k_x + Q_y\cos k_y),
\eqno(9)
$$
while bosonic Hamiltonian is diagonalised via a
Boguilobov transformation. This must be handled with care because
of the possibility of a gapless spectrum and hence Bose-Einstein
condensation at zero temperature. For the wavevectors at which
Bose-Einstein condensation {\it does not occur} a non-unitary
transformation is employed\refto{ring} which yields the
eigenfunctions:   $$
\left( \alpha_k^\dagger, \beta_k \right) =
\left(
b_{k\uparrow}^\dagger,
b_{-k\downarrow}
\right)
\left( \matrix{
{\sqrt{\xi +1 \over 2}}e^{i\th/2} & {\sqrt{\xi -1 \over
2}}e^{i\th/2} \cr
{\sqrt{\xi -1 \over 2}}e^{-i\th/2} & {\sqrt{\xi +1 \over
2}}e^{-i\th/2} \cr
}
\right)
\eqno(10a)
$$
and corresponding eigenvalues,
$$
\eps^b_k = \pm \sqrt{(\lambda_b+M_k)^2 -|\gamma_k|^2},
\eqno(10b)
$$
where $\xi_k = (\lam_b+M_k)/\eps_k^b$ and $\th$ is the phase of
$\gamma$. The operators $\alp_k^\dg$ and $\beta_k^\dg $ obey Bosonic
commutation relations.

If there exist wave vectors where the spectrum is gapless, however, a
unitary transformation is required to diagonalise the Hamiltonian
(6b). The eigenfunctions are then $$
\left( \zeta_k^\dagger, \eta_k^\dagger \right)=
\left(b_{k\uparrow}^\dagger,b_{-k\downarrow}\right)
\left( \matrix{
{ e^{i\th/2} \over {\sqrt 2}} & { e^{i\th/2} \over {\sqrt 2}} \cr
{ e^{-i\th/2} \over {\sqrt 2}} &{ e^{-i\th/2} \over {\sqrt 2}} \cr
}
\right),
\eqno(11)
$$
with eigenvalues of $2\lam_b$ and $0$. These operators obey the
commutation relations,
$[\zeta_k,\zeta_k^\dagger]=[\eta_k,\eta_k^\dagger]=0$ and
$[\zeta_k,\eta_k^\dagger] =[\eta_k,\zeta_k^\dagger]=1$. The
occupation of the Bose condensate is given by,
$n_0=\sum_k<\eta_k^\dagger \eta_k>$.
As usual, the occurence of Bose Einstein condensation means that the
bosonic chemical potential is specified. In particular,
$$
\lam_b = 2\sum_{\alp=x,y}\sqrt{ J_\del^2D_\alp ^2/2
+(F_\alp t+Q_\alp J_\del/4)^2}
\eqno(12a)
$$
with the condensed modes being at
$$K^0_\alp=\tan^{-1}\left({- J_\delta D_\alp /\sqrt{2} \over F_\alp t
+ Q_\alp J_\del/4} \right). \eqno(12b)
$$
The physical interpretation of a gapless spectrum and Bose-Einstein
condensation is that it signifies long range magnetic order. The
condensed bosons correspond to the classical spin component, while
the normal bosons represent the zero point quantum mechanical
fluctuations. The pitch of the classical spiral is $2(K_x^0,K_y^0)$,
which is readily related to the classical angle, $\theta$, eqn. (5).
Absence of a Bose-Einstein condensate implies that the quantum
mechanical fluctuations are dominate, leading to a disordered spin
liquid with no long range correlations.

Having determined the eigensolutions of (3) the
self-consistent equations for the order parameters are obtained.
These are, at zero temperature:
$$
2-\delta = {1 \over N} \sum_k \left[ \left({\lambda_b + M_k \over
\eps_k^b} \right)(1 - \delta_{kK_0}) + n_0\delta_{kK_0} \right],
\eqno(13a)
$$
$$
D_\alp = {J_\del \over N} \sum_k \sin k_\alp  \left[ {(1 -
\delta_{kK_0})  \over \eps_k^b}
+ { n_0 \delta_{kK_0} \over |\lambda_b + M_k |} \right]
\sum_{\beta=x,y}  D_\beta \sin k_\beta,
\eqno(13b)
$$
$$
Q_\alp={1 \over N} \sum_k \cos k_\alp \left[  \left({\lambda_b + M_k
\over \eps_k^b} \right)(1 - \delta_{kK_0}) + n_0\delta_{kK_0}
\right], \eqno(13c)
$$
$$
F_\alp = {1 \over N}\sum_{k \leq k_f} \cos k_\alp
\eqno(13d)
$$
and
$$
\delta = {1 \over N} \sum_{k \leq k_f}.
\eqno(13e)
$$
If Bose-Einstein condensation does exist (13a) can be regarded as a
defining $n_0$ with $\lam_b$ defined by (12). Otherwise, it defines
$\lam_b$. (13e) is a sum up to the Fermi wavevector for the
fermions, and hence defines $\lam_f$.
The equations (13) are solved numerically, with the ground state
mean field energy  per site being given by\refto{auerbach2},
$$
E^{MF}=-t(Q_xF_x+Q_yF_y) - J_\del(D_x^2+D_y^2)
+J_\del(Q_x^2+Q_y^2)/4. \eqno(14)
$$
The resulting phase diagram is
discussed in the next section.

\subhead{3. The Phase Diagram}

Figure (1) shows the diagram associated with the
mean field phases which minimise the energy. For zero doping
the 2-D Heisenberg anti-ferromagnet is stable. As holes are doped
into the anti-ferromagnet the (1,1) spiral develops. However, we
note that the (1,0) spiral is also very close in energy if
isotropic $D$ {\it is not assumed}. For large $t/J$ there is a rapid
decrease in the classical nearest neighbour angle,
$\theta_{ij}=\cos^{-1}\left(\^{\Omega_i}.\^{\Omega_j}\right)$, while
for small $t/J$ the spins remain nearly anti-ferromagnetically
aligned for quite large doping. Figure (2) illustrates the nearest
neighbour  classical angle as a function of doping for $t/J=2.0$.
For small deviations from N\'{e}el order the classical angle is
approximately given by: $$
\pi - \theta_{ij} \simeq  {\pi \over 2} {t \over J}\del.
$$
Neutron scattering experiments would therefore give Bragg peaks at
Q vectors $ \pi(1 \pm t\del/\sqrt{2}J,1 \pm
t\del/\sqrt{2}J)$, which, although being in the wrong
direction, has a magnitude in close agreement with experimental Q
vectors of $\pi (1 \pm 2\del, 1)$ and $\pi(1, 1 \pm
2\del)$ for La$_{2-x}$Sr$_x$CuO$_4$\refto{hayden}.

By allowing the order parameters to be anisotropic we find new
energetically favourable phases. In particular, if one of the
components of $D$ is chosen to be zero, then the component of $Q$
in the orthogonal direction is zero. This is the collinear phase in
which the spins are aligned anti-ferromagnetically in one direction
and ferromagnetically in the other. As the doping in increased there
is a first order transition to this collinear phase from the (1,1)
spiral. The collinear phase is stable for quite high doping before a
further first order phase transition to the ferromagnet phase
occurs.

Chakraborty {\it et al.}\refto{kane} introduced the so-called `flux
phase' which breaks time reversal invariance: $D_x = iD_y$. They
showed that this phase is more stable than the Nagaoka ferromagnet.
Unfortunately, we were unable to find reliable mean field solutions
of this phase to test it against the collinear phase. However, the
dashed line in figure (1) indicates the phase boundary line between
the (1,1) spiral and the flux phase as taken from
ref[6]\refto{footnote}. The position of this line indicates that
the collinear phase is stable with respect to the (1,1) spiral in
parameter ranges where the flux phase is not stable.

In most regions of the phase diagram Bose-Einstein condensation
occurs, which implies long range order. Figure (2) shows the
fraction of condensed ({\it i.e.} classical) spins as a function of
doping, defined by $2n_0/N +\del$. As the doping is increased, and
the spins become more ferromagnetically aligned, the classical
component of the spins increases so that at the ferromagnetic regime
there are no zero point fluctuations. Notice that there are
discontinuities in the condensed fraction at the phase boundaries.
Bose-Einstein condensation is, however, an artefact of the mean field
calculation. If fluctuations in the phase of $<\D_{ij}>$, which
arise from the holes hopping across bonds, are taken into account
then long range phase coherence would presumably be lost.

\subhead{4. Conclusions}

We have extended the Schwinger Boson mean field theories of the
`t-J' model by allowing for anisotropic order parameters. This has
two effects. First, the (1,1) and (1,0) spiral phases are now very
close in energy. The inclusion of
 arbitrarily weak intra-sublattice coupling may therefore stabilise
the (1,0) spiral with respect to the (1,1) spiral, in agreement
with the experimental observations. Second, a collinear phase, in
which the spins are anti-ferromagnetically aligned in one direction
and ferromagnetically aligned in the other, is found to be stable
over a significant range of the phase diagram.

\subhead{Acknowledgements} We thank the SERC
(United Kindgom) for financial support (grant ref. GR/F 75445).

\vfill
\eject

\head{References}

\refis{hayden} S-W Cheong, G Aeppli, T E Mason, H Mook, S M Hayden,
P C Canfield, Z Fisk, K N Clausen and J L Martinez, Phys. Rev.
Lett., 67, 1791 (1991).

\refis{shraiman} B I Shraiman and  E D Siggia, Phys. Rev. Lett, 62,
1564 (1989).

\refis{barford} W Barford and J P Lu, Phys. Rev. B, 43, 3540 (1991).

\refis{lu} J P Lu, Q M Si, J H Kim and K Levin, Phys. Rev. Lett.,
65, 2466 (1990).

\refis{yoshioka} D Yoshioka, J. Phys. Soc Japan, 58, 1516 (1989),
C. Jayaprakash, H R Krishnamurthy and S Sarker, Phys. Rev. B, 40,
2610 (1989), C L Kane, P A Lee, T K Ng, B Chakraborty and N Read,
{\it ibid}, 41, 2653 (1990).

\refis{kane} B Chakraborty, N Read, C Kane and P A Lee, Phys. Rev.
B, 42, 4819 (1990).

\refis{gan} J Gan and F Mila, Phys. Rev. B, 44, 12624 (1991).

\refis{auerbach} For details of the Hubbard-Stratonovich
transformations see  A Auerbach and B E Larson, Phys. Rev. B, 43,
7800 (1991).

\refis{ring} P Ring and P Schuck, {\it The Nuclear Many-body
Problem}, (Springer Verlag, New York 1980).

\refis{footnote} This data was taken from ref[6] assuming that the
J used in their calculation is the mean field value of J,
{\it i.e.}, $J(1 - \del)^2$.

\refis{auerbach2} {{\it ibid.}\refto{auerbach}}

\endreferences

\vfill\eject

\head{Figure Captions}

Figure 1.

Mean field phase diagram of the `t-J' model. The dashed line
indicates the phase boundary line between the (1,1) spiral and the
flux phase as taken from ref[6].

Figure 2.

The nearest neighbour classical angle, $\theta_{ij}/\pi$, (dashed
line) and the fraction of condensed (classical) spins (solid line)
versus doping for $t/J =2.0$. Note that in the collinear region
$\theta_{ij}$ is  $\pi$ in one direction and $0$ in the orthogonal
direction.

\endit \bye